\documentclass[12pt,a4paper]{article}
\usepackage[utf8]{inputenc}
\usepackage{a4wide}
\usepackage{graphicx}
\usepackage{amsmath}
\usepackage{dcolumn}
\allowdisplaybreaks[1]
\usepackage{hyperref}

\newcommand{\hPi}{\hat\Pi}
\newcommand{\pPi}{\widetilde\Pi}
\newcommand{\pip}{\pi^{+}}

\newcommand{\pio}{\pi^{0}}
\newcommand{\bigG}{\mathcal{G}}

\begin{document}

\begin{titlepage}
\begin{flushright}
LU TP 16-49\\
arXiv:1609.01573 [hep-lat]\\
Revised November 2016
\end{flushright}
\vfill
\begin{center}
{\Large\bf Connected, Disconnected and Strange Quark\\[3mm]Contributions to HVP}
\vfill
{\bf Johan Bijnens and Johan Relefors}\\[0.3cm]
{Department of Astronomy and Theoretical Physics, Lund University,\\
S\"olvegatan 14A, SE 223-62 Lund, Sweden}
\end{center}
\vfill
\begin{abstract}
We calculate all neutral vector two-point functions in
Chiral Perturbation Theory (ChPT)
to two-loop order and use these to estimate the ratio of disconnected to
connected contributions as well as contributions involving the strange quark.
We extend the ratio of $-1/10$ derived earlier in 
two flavour ChPT at one-loop order to a large part of the higher order
contributions and discuss corrections to it.
Our final estimate of the ratio disconnected to connected is negative
and a few \% in magnitude.
\end{abstract}
\vfill
\end{titlepage}

\section{Introduction}

The muon anomalous magnetic moment is one of the most precisely measured
quantities around. The measurement \cite{Bennett:2006fi} differs from
the standard model prediction by about 3 to 4 sigma depending on precisely
which theory predictions are taken. A review
is~\cite{Jegerlehner:2009ry} and talks on the present situation
can be found in~\cite{Proceedings:2016bng}. The main part of the theoretical
error at present is from the lowest-order hadronic vacuum polarization (HVP).
This contribution can be determined from experiment or can
be computed using lattice QCD~\cite{Blum:2002ii}.
An overview of the present situation in lattice QCD calculations
is given by~\cite{WittigLattice2016}.

\begin{figure}
\begin{center}
\includegraphics[width=8cm]{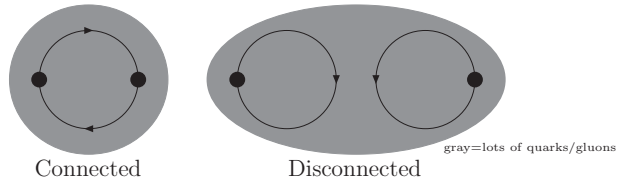}
\end{center}
\caption{Connected (left) and disconnected (right) diagram for the two-point
 vector function. The lines are valence quark lines in a sea of quarks and
 gluons.}
\label{fig:condis}
\end{figure}

The underlying object that needs to be calculated is the
two-point function of electromagnetic currents as defined
in (\ref{deftwopoint}). The contribution to $a_\mu = (g-2)/2$ is given
by the integral in (\ref{amuint}).
There are a number of different contributions to the two-point function
of electromagnetic currents  that need to be measured on the lattice.
First, if we only consider the light up and down quarks, there
are connected and disconnected contributions depicted schematically in
Fig.~\ref{fig:condis}.
If we add the strange quark to the electromagnetic currents then there
are contributions with the strange electromagnetic current
in both points and the mixed up-down and strange case.
In this paper we provide estimates of all contributions at low energies
using Chiral Perturbation Theory (ChPT).

The disconnected light quark contribution
has been studied at one-loop order in Ref.~\cite{DellaMorte:2010aq}
using partially quenched (PQChPT). They found that the ratio
in the subtracted form factors, as defined in (\ref{eq:subtraction}), is $-1/2$
in the case of valence quarks of a single mass and two degenerate sea quarks.
They also found that adding the strange quark did not change the ratio much.
Here we give an argument explaining the factor of $-1/2$ and
extend their analysis to order $p^6$. We also present estimates
for the contributions from the strange electromagnetic current. 

The finite volume,  partially quenched and twisted boundary conditions
extensions to two loop order will be presented in \cite{JBJR}.

In Sect.~\ref{twopoint} we give the definitions of the two-point functions
and currents we use. Sec.~\ref{chpt} discusses ChPT and the extra terms
and low-energy-constants (LECs) needed for a singlet vector current.
Our main analytical results, the two-loop order ChPT expressions for all
needed vector two-point functions are in Sect.~\ref{chptresults}.
Section~\ref{disconnected} uses the observation given in Sect.~\ref{chpt} of the
absence of singlet vector couplings to mesons until ChPT order $p^6$
to show for which contributions the ratio $-1/2$ is valid.
Numerical results need an estimate of the LECs involved, both old and new.
This is done in Sect.~\ref{estimatedisconnected} and applied there to
the light connected and disconnected part. Because of the presence of the LECs
we find a total disconnected contribution of opposite sign and size a few \% of
the connected contribution. The same type of estimates are then used for the
strange quark contribution in Sect.~\ref{estimatestrange}. Here
we find a very strong cancellation between $p^4$ and $p^6$ contributions,
leaving the LEC part dominating strongly.
A comparison with a number of lattice results is done in
Sect.~\ref{latticecomparison}. We find a reasonable agreement in some cases.
Our conclusions are summarized in Sect.~\ref{conclusions}.

\section{The vector two-point function}
\label{twopoint}

We define the two-point vector function as
\begin{align}
\label{deftwopoint}
\Pi^{\mu\nu}_{ab} = i \int d^4x e^{iq\cdot x}\left<T(j^\mu_a(x)j^{\nu\dagger}_b(0))\right>
\end{align}
where the labels $a,b$ specify the involved currents. We label the currents
as
\begin{align}
j_{\pip}^\mu =\, &\bar d \gamma^\mu u\,, &
j_{U}^\mu =\,& \bar u \gamma^\mu u\,, &
j_{D}^\mu =\,& \bar d \gamma^\mu d\,, 
\nonumber\\
j_{S}^\mu =\,& \bar s \gamma^\mu s \,, & 
j_{EM}^\mu =\,& 
  \frac{2}{3}j_{U}^\mu - \frac{1}{3}j_{D}^\mu - \frac{1}{3}j_{S}^\mu\,, &
j_{EM2}^\mu =\,& \frac{2}{3}j_{U}^\mu - \frac{1}{3}j_{D}^\mu\,,
\nonumber\\
j_{\pio}^\mu =\, &
   \frac{1}{\sqrt{2}}\left(j_U^\mu-j_D^\mu\right)\,,&
j_{I2}^\mu =\, &
   \frac{1}{\sqrt{2}}\left(j_U^\mu+j_D^\mu\right) \,,&
j_{I3}^\mu =\, &
   \frac{1}{\sqrt{3}}\left(j_U^\mu+j_D^\mu+j_S^\mu\right) \,.
\end{align}
The divergence of the vector current is given by
\begin{align}
  \partial_\mu \bar q_i\gamma^\mu q_j = i(m_i-m_j)\bar q_i q_j\,,
\end{align}
which means that any current involving equal mass quark and anti-quark
is conserved. Assuming isospin for the $\pip$ current,
Lorentz invariance then implies that we can parametrize the vector two-point
functions given above as
\begin{align}
  \Pi^{\mu\nu}_{ab}(q) = (q^\mu q^\nu - q^2g^{\mu\nu})\Pi_{ab}(q^2).
\end{align}
We also define the subtracted quantity
\begin{align}
  \label{eq:subtraction}
  \hPi_{ab}(q^2) = \Pi_{ab}(q^2) - \Pi_{ab}(0)\,.
\end{align}
For simplicity we also use $\Pi_a=\Pi_{aa}$ and $\hPi_a=\hPi_{aa}$

In this paper we work in the isospin limit. This immediately leads to a number
of relations
\begin{equation}
\label{isorelations}
\Pi_{\pip} = \Pi_{\pio}\,,\qquad \Pi_U = \Pi_D\,,\qquad \Pi_{US} = \Pi_{DS}\,. 
\end{equation}
With those one can derive
\begin{align}
\label{defPiEM}
\Pi_{EM} =\,& \frac{5}{9}\Pi_U+\frac{1}{9}\Pi_S-\frac{4}{9}\Pi_{UD}
         -\frac{2}{9}\Pi_{US}\,,
\nonumber\\
\Pi_{EM2} =\,& \frac{5}{9}\Pi_U-\frac{4}{9}\Pi_{UD}\,.
\end{align}

The two-point functions $\Pi$ are themselves not directly observable.
However, the vector current two-point function in QCD satisfies
a once subtracted dispersion relation
\begin{equation}
\hPi(q^2) = \Pi(q^2)-\Pi(0) =  q^2\int_{threshold}^\infty ds \frac{1}{s(s-q^2)}\frac{1}{\pi}
\mathrm{Im}\Pi(s)\,.  
\end{equation}
The imaginary part can be measured in hadron production if there exists an
external vector boson like $W^\pm$ or the photon coupling to the current.
Thus $\hPi(q^2)$ is an observable, but not $\Pi(0)$. $\Pi(0)$ depends
on the precise definitions used in regularizing the product of two currents
in the same space-time point.
The two-point functions for the electromagnetic current can be determined
in $e^+e^-$ collisions and $\Pi_{\pip}$ in $\tau$-decays.

One main use of $\hPi$ is the determination of the lowest order HVP
part of the muon anomalous magnetic moment via the integral over the
electromagnetic two-point function\footnote{The version mentioned here
comes from \cite{Blum:2002ii}
but the result essentially goes back to \cite{Bouchiat,Durand:1962zzb}}
\begin{align}
\label{amuint}
a_\mu^{LOHVP} =\,& 4\alpha^2\int_0^\infty dQ^2 \hPi_{EM}(-Q^2) g(Q^2)\,,
\nonumber\\
g(Q^2) =\,&\frac{-16 m_\mu^4}{Q^6\left(1+\sqrt{1+4 m_\mu^2/Q^2}\right)^4\sqrt{1+4 m_\mu^2/Q^2}}\,.
\end{align}

\section{Chiral perturbation theory and the singlet current}
\label{chpt}

ChPT describes low-energy QCD as an expansion in masses and momenta
\cite{Weinberg1,GL1,GL2}. The dynamical degrees of freedom are the
pseudo-Goldstone bosons (GB) from the spontaneous breaking of the left- and
right-handed flavor symmetry to the vector subgroup, 
$SU(3)_L\times SU(3)_R\rightarrow SU(3)_V$. The GB can be parameterized in
the $SU(3)$ matrix
\begin{align}
  U = e^{i\sqrt{2}M/F_0}
\quad
\text{with}
\quad
  M =
  \left(\begin{matrix}
    \frac{1}{\sqrt{2}}\pi^0+\frac{1}{\sqrt{6}}\eta & \pi^+ & K^+ \\
    \pi^- & -\frac{1}{\sqrt{2}}\pi^0+\frac{1}{\sqrt{6}}\eta & K^0 \\
    K^- & \bar{K}^0 & -\frac{2}{\sqrt{6}}\eta
  \end{matrix}\right).
\end{align}
or with the $2\times 2$ matrix with only the pions in the case of two-flavours.
The Lagrangians, as well as the divergences, are known at
order $p^2$ (LO), $p^4$ (NLO) and $p^6$(NNLO)
in the ChPT counting \cite{GL1,GL2,BCE1,BCE2}.
However, the vector currents defined
in Sect.~\ref{twopoint} contain also a singlet component and the Lagrangians
including only this extension are not known. There is work when extending
the symmetry to including the singlet GB as well as singlet vector and
axial-vector currents at $p^4$ \cite{HerreraSiklody:1996pm}
and $p^6$ \cite{Jiang:2014via}. However this contains very many more
terms than we need. If we only add the singlet vector current, in addition
to simply extending the external vector field to include the singlet part,
there are two extra terms relevant at order $p^4$:
\begin{equation}
H_3\left(\left<F_{L\mu\nu}\right>\left<F_L^{\mu\nu}\right>
        +\left<F_{R\mu\nu}\right>\left<F_R^{\mu\nu}\right>\right)
+H_4\left<F_{R\mu\nu}\right>\left<F_L^{\mu\nu}\right>\,.
\end{equation}
Since we are only interested in two-point functions of vector currents
these will always appear in the combination $2 H_3+H_4$.
For the two-flavour case we get $H_3\to h_4$ and $H_4\to h_5$ but otherwise similar terms.

It should be noted that none of the terms in the extended $p^4$ Lagrangian
contains
couplings of the singlet vector-field to the GB. The singlet appearing in
commutators vanishes and the terms involving field strengths
vanish, except for the combinations above which do not contain GB fields.

At order $p^6$ there are many more terms, specifically there are terms
appearing that contain interactions of the singlet vector field with the GBs.
Two  examples are
\begin{equation}
\left<F_{R\mu\nu}\chi F_L^{\mu\nu} U^\dagger\right>
+\left<F_{L\mu\nu}\chi^\dagger F_R^{\mu\nu} U\right>\,,\qquad
\left<F_{L\mu\nu}+F_{R\mu\nu}\right>\left<\left(\chi U^\dagger+U\chi^\dagger\right)
 D^\mu U D^\nu U^\dagger\right>\,.
\end{equation}
The extra terms that contribute to the vector two-point function at order $p^6$
always contain two field strengths and the extra $p^2$ needed
can come from either
two derivatives or quark masses. Setting all GB fields to zero,
the only possible extra terms have a structure with $F_{V\mu\nu}$
the vector-field field strength and $\bar\chi$ the quark mass part of
$\chi$. This leads to the possible terms
\begin{equation}
D_1\left<F_{V\mu\nu}\right>\left<F_V^{\mu\nu}\bar\chi\right> 
+D_2\left<F_{V\mu\nu}\right>\left<F_V^{\mu\nu}\right>\left<\bar\chi\right> 
+D_3
\left<\partial_\rho F_{V\mu\nu}\right>\left<\partial^\rho F_V^{\mu\nu}\right>
\end{equation}
The $D_i$ are linear combinations of a number of LECs in the Lagrangian
and one can check that they are all independent by writing down a few fully
chiral invariant terms.
A similar set with $D_i\to d_i$ exists for the two-flavour case.

There is a coupling of the singlet vector current to the GBs already at
order $p^4$ via the Wess--Zumino--Witten (WZW) term. However, due to the presence of
$\epsilon^{\mu\nu\alpha\beta}$ we need an even number of insertions
of the WZW term or higher order terms from the odd-intrinsic-parity sector
to get a contribution to the vector two-point functions.

\section{ChPT results up to two-loop order}
\label{chptresults}

The vector two-point functions for neutral non-singlet currents
were calculated in \cite{Golowich:1995kd,Amoros:1999dp}. We have reproduced
their results and added the parts coming from the singlet currents.

The expressions for the two-point functions are most simply expressed
in terms of the function
\begin{align}
\bigG(m^2,q^2) \equiv \frac{1}{q^2}\left(\overline B_{22}(m^2,m^2,q^2)
                                        -\frac{1}{2}\overline A(m^2)\right)
\end{align}
The one-loop integrals here are defined in many places,
see e.g. \cite{Amoros:1999dp}. The explicit expression is
\begin{align}
\bigG(m^2,q^2) =\,&\frac{1}{16\pi^2}\left[
    \frac{1}{36}+\frac{1}{12}\log\frac{m^2}{\mu^2}
    +\frac{q^2-4m^2}{12}\int_0^1 dx \log\left(1-x(1-x)\frac{q^2}{m^2}\right)
                                    \right]
\nonumber\\
=\,&\frac{1}{16\pi^2}\left(
    \frac{1}{12}+\frac{1}{12}\log\frac{m^2}{\mu^2}
    -\frac{q^2}{12m^2}-\frac{q^4}{1680m^4}+\cdots
                                    \right)
\end{align}
We also need
\begin{equation}
\overline A(m^2) = -\frac{m^2}{16\pi^2}\log\frac{m^2}{\mu^2}\,.
\end{equation}
$\mu$ is the ChPT subtraction scale. We always work in the
isospin limit. The expressions we give are in the three flavour case
with physical masses. We will quote the corresponding results
with lowest order masses in \cite{JBJR}.

The two-point functions only start at $p^4$. We therefore write the result as
\begin{equation}
\Pi = \Pi^{(4)} + \Pi^{(6)}+\cdots
\end{equation}
in the chiral expansion.
The $p^4$ results are
\begin{align}
\label{resultp4}
\Pi_{\pi^+}^{(4)} =\,& -8 \bigG(m_\pi^2,q^2)-4 \bigG(m_K^2,q^2)-4(L_{10}^r+2H_1^r)\,,
\nonumber\\
\Pi_U^{(4)} =\,&  -4 \bigG(m_\pi^2,q^2)-4 \bigG(m_K^2,q^2)
                     -4(L_{10}^r+2 H_1^r+2 H_3^r+H_4^r)\,,
\nonumber\\
\Pi_S^{(4)} =\,&  -8 \bigG(m_K^2,q^2)
                     -4(L_{10}^r+2 H_1^r+2 H_3^r+H_4^r)\,,
\nonumber\\
\Pi_{UD}^{(4)} =\,&   4 \bigG(m_\pi^2,q^2)
                     -4(2 H_3^r+H_4^r)\,,
\nonumber\\
\Pi_{US}^{(4)} =\,&   4 \bigG(m_K^2,q^2)
                     -4(2 H_3^r+H_4^r)\,,
\nonumber\\
\Pi_{EM}^{(4)} =\,& - 4\bigG(m_\pi^2,q^2) - 4\bigG(m_K^2,q^2) -\frac{8}{3}(L_{10}^r+2 H_1^r)\,.
\end{align}
The obvious relations visible for the $\bigG$ terms
will be discussed in Sect.~\ref{disconnected}.
This result agrees with \cite{DellaMorte:2010aq} when the appropriate limits
are taken.

The results at $p^6$ are somewhat longer but still fairly short.
\begin{align}
\label{resultp6}
F_\pi^2\Pi_{\pi^+}^{(6)} =\,&
4 q^2\left(2\bigG(m_\pi^2,q^2)+\bigG(m_K^2,q^2)\right)^2
-16 q^2 L_9^r \left(2\bigG(m_\pi^2,q^2)+\bigG(m_K^2,q^2)\right)
\nonumber\\&
-8 (L_9^r+L_{10}^r)\left(2\overline A(m_\pi^2)+\overline A(m_K^2)\right)
-32m_\pi^2C_{61}^r-32(m_\pi^2+2m_K^2)C_{62}^r-8 q^2 C_{93}^r\,,
\nonumber\\
F_\pi^2\Pi_U^{(6)} =\,& 8q^2\bigG(m_\pi^2,q^2)^2+ 8q^2\bigG(m_\pi^2,q^2)\bigG(m_K^2,q^2) 
               +8q^2\bigG(m_K^2,q^2)^2
\nonumber\\ &
-16 q^2 L_9^r \left(\bigG(m_\pi^2,q^2)+\bigG(m_K^2,q^2)\right)
-8 (L_9^r+L_{10}^r)\left(\overline A(m_\pi^2)+\overline A(m_K^2)\right)
\nonumber\\ &
-32m_\pi^2C_{61}^r-32(m_\pi^2+2m_K^2)C_{62}^r-8 q^2 C_{93}^r
-4 m_\pi^2 D_1^r-4(m_\pi^2+2m_K^2)D_2^r-4q^2D_3^r\,,
\nonumber\\
F_\pi^2\Pi_S^{(6)} =\,& 24 q^2\bigG(m_K^2,q^2)^2
-32 q^2 L_9^r \bigG(m_K^2,q^2)
-16 (L_9^r+L_{10}^r)\overline A(m_K^2)
\nonumber\\ &
-32(2m_K^2-m_\pi^2)C_{61}^r-32(m_\pi^2+2m_K^2)C_{62}^r-8 q^2 C_{93}^r
\nonumber\\&
-4 (2 m_K^2-m_\pi^2) D_1^r-4(m_\pi^2+2m_K^2)D_2^r-4q^2D_3^r\,,
\nonumber\\
F_\pi^2\Pi_{UD}^{(6)} =\,& -8q^2\bigG(m_\pi^2,q^2)^2
                 -8q^2\bigG(m_\pi^2,q^2)\bigG(m_K^2,q^2) 
               +4q^2\bigG(m_K^2,q^2)^2
\nonumber\\ &
              +16 q^2 L_9^r\bigG(m_\pi^2,q^2)
                 +8(L_9^r+L_{10}^r)\overline A(m_\pi^2)     
-4 m_\pi^2 D_1^r-4(m_\pi^2+2m_K^2)D_2^r-4q^2D_3^r\,,
\nonumber\\
F_\pi^2\Pi_{US}^{(6)} =\,& -12q^2 \bigG(m_K^2,q^2)^2 
                 +16 q^2 L_9^r\bigG(m_K^2,q^2)
                 +8(L_9^r+L_{10}^r)\overline A(m_K^2)
\nonumber\\&
-4  m_K^2 D_1^r-4(m_\pi^2+2m_K^2)D_2^r-4q^2D_3^r\,.
\end{align}

For the two-flavour case the results can be derived from the above.
First, only keep the integral terms with $m_\pi^2$, second
replace $L_9$ by $-(1/2) l_6^r$,
$L_{10}^r+2H_1^r$ by $-4 h_2^r$ and  $L_{10}^r$ by $l_5^r$.
In addition there are also extra counterterms for the singlet current
appearing. The results are
\begin{align}
\label{resultNf2}
\Pi_{\pi^+}^{(4)} =\,& -8 \bigG(m_\pi^2,q^2)+16 h_2^r\,,
\nonumber\\
\Pi_U^{(4)} =\,&  -4 \bigG(m_\pi^2,q^2)+16 h_2^r-4(2 h_4^r+h_5^r)\,,
\nonumber\\
\Pi_{UD}^{(4)} =\,&   4 \bigG(m_\pi^2,q^2)
                     -4(2 h_4^r+h_5^r)\,,
\nonumber\\
\Pi_{EM}^{(4)} =\,& - 4\bigG(m_\pi^2,q^2)  +\frac{32}{3}h_2^r-\frac{4}{9}(2 h_4^r+h_5^r)\,,
\nonumber\\
F_\pi^2\Pi_{\pi^+}^{(6)} =\,&
16 q^2\bigG(m_\pi^2,q^2)^2
+16 q^2 l_6^r \bigG(m_\pi^2,q^2)
-8 (2l_5^r-l_6^r)\overline A(m_\pi^2)
-32m_\pi^2c_{34}^r-8 q^2 c_{56}^r\,,
\nonumber\\
F_\pi^2\Pi_U^{(6)} =\,& 8q^2\bigG(m_\pi^2,q^2)^2
+8 q^2 l_6^r \bigG(m_\pi^2,q^2)
-4 (2 l_5^r-l_6^r)\overline A(m_\pi^2)
\nonumber\\ &
-32m_\pi^2c_{34}^r-8 q^2 c_{56}^r
-4 m_\pi^2 (d_1^r+2d_2^r)-4q^2d_3^r\,,
\nonumber\\
F_\pi^2\Pi_{UD}^{(6)} =\,& -8q^2\bigG(m_\pi^2,q^2)^2
 -8 q^2 l_6^r\bigG(m_\pi^2,q^2)
 +4(2l_5^r-l_6^r)\overline A(m_\pi^2)     
-4 m_\pi^2(d_1^r+2d_2^r)-4q^2d_3^r\,.
\end{align}

\section{Connected versus disconnected contributions}
\label{disconnected}

If we look at the flavour content of the two-point functions in the
isospin limit, it is clear that $\Pi_{\pip}$ only contains connected
contributions while $\Pi_{UD}$ only contains disconnected contributions.
This is derived by thinking of which quark contractions can contribute
as shown in Fig.~\ref{fig:condis}.
In the same way $\Pi_U$ contains both with
\begin{equation}
\label{relPiU}
\Pi_U = \Pi_{\pip}+\Pi_{UD}\,.
\end{equation}
Inspection of all the results in Sect.~\ref{chptresults} shows
that (\ref{relPiU}) is satisfied. From (\ref{defPiEM}) we thus obtain
\begin{equation}
\label{PiEM2}
\Pi_{EM2} = \frac{5}{9}\Pi_{\pip}+\frac{1}{9}\Pi_{UD}\,,
\end{equation}
and
\begin{equation}
\label{PiEM}
\Pi_{EM} = \frac{5}{9}\Pi_{\pip}+\frac{1}{9}\Pi_{UD}-\frac{2}{9}\Pi_{US}+\frac{1}{9}\Pi_S\,.
\end{equation}
$\Pi_{US}$ is fully disconnected while $\Pi_S$ has both connected
and disconnected parts.

\subsection{Two-flavour and isospin arguments}

In \cite{DellaMorte:2010aq}, they found, using NLO two-flavour ChPT
in the isospin limit, that
\begin{align}
  \label{eq:DellaMorteSU2}
  \frac{\hPi_{EM2}^{Disc}}{\hPi_{EM2}^{conn}} = -\frac{1}{10}\,.
\end{align}
They also calculated corrections to this ratio due to the inclusion of strange
quarks. Their result is in our terms expressed via
\begin{equation}
\label{relNf2}
\frac{\hPi_{UD}^{(4)}}{\hPi_{\pip}^{(4)}} = -\frac{1}{2}
\end{equation}
which is clearly satisfied for the results shown in (\ref{resultNf2}).
Note that $\Pi(0)$, via the part coming from the LECs,
does not satisfy a similar
relation due to the extra terms possible for the singlet current.
Inspection of (\ref{resultNf2}) shows that the loop part at order $p^6$
also satisfies (\ref{relNf2}) but due to the part of the LECs, the relation
is no longer satisfied even for the subtracted functions $\hPi$.

The relation (\ref{relNf2}) can be derived in more general way.
As noted in Sect.~\ref{chpt} the singlet current $j_{I2}^\mu$ only couples
to GBs at order $p^6$ or at order $p^4$ via the WZW term and we need at least
two of the latter for the vector two-point function. For the contributions
where those couplings are not present,
denoted by a tilde, we get
\begin{equation}
\label{relprimeNf2}
\pPi_{U(U+D)} = \pPi_U+\pPi_{UD} = 0\,,
\end{equation}
The relation (\ref{relprimeNf2}) if written for $\hPi$ has corrections
at order $p^8$.
Eq. (\ref{relprimeNf2})
together with (\ref{relPiU}) immediately leads to (\ref{relNf2})
but for many more contributions. 
The ratio of disconnected to connected
is $-1/2$ for all loop-diagrams only involving vertices from the lowest-order
Lagrangian or from the normal NLO Lagrangian. So the ratio is true for a large
part of all higher order loop diagrams and corrections start appearing
only in loop diagrams at order $p^8$ with one insertion from the
$p^6$-Lagrangian or at $p^{10}$ with two insertions of a WZW vertex.
The argument includes diagrams with four or more pions.

Using the isospin relations we can derive that
\begin{equation}
\label{reliso}
\Pi_{UD} = \frac{1}{2}\left(\Pi_{I2}-\Pi_{\pi^0}\right)
\end{equation}
Looking at (\ref{reliso}), one can see that the ratio $(-1/2)$ is exact for all
contributions with isospin $I=1$ and only broken due to $I=0$ contributions.
This can be used as well to estimate the size of the ratio,
see below and \cite{Francis:2013qna,Chakraborty:2015ugp}. A corollary is
that two-pion intermediate state contributions obey (\ref{relNf2}) to all
orders. 

The contributions to order $p^6$ for $\hPi$ satisfy the
relation (\ref{relprimeNf2}) up to the
LEC contributions. Using resonance saturation, the LECs can be
estimated from $\rho$ and $\omega$ exchange. In the large $N_c$ limit
that combination will only contribute to the connected contribution.
Since the $\rho$-$\omega$ mass splitting and coupling differences are rather
small, we expect that the disconnected contribution from this source will
be rather small. This will lower the ratio of disconnected
to connected contributions compared
to (\ref{relNf2}).

In \cite{Francis:2013qna} it was also noticed that the ratio of $-1/2$
is valid for all two-pion intermediate states in the isospin limit.
They used the slow turn-on of the three-pion channel where the singlet
starts contributing to argue for the validity of the one-loop estimate.
That slow turn-on follows from the three-pion contribution being $p^{10}$
in our way of looking at it.   
In \cite{Chakraborty:2015ugp} the difference between $\rho$ and
$\omega$ measured masses and couplings were used to obtain an estimate of
the disconnected contribution of about $-1\%$.
We consider that contribution to be within the
error of our estimate given in Sect.~\ref{estimatedisconnected}.

\subsection{Three flavour arguments}

It was already noted in \cite{DellaMorte:2010aq} that kaon loops violate
the relation (\ref{relNf2}) in NLO three-flavour ChPT and the same is rather
visible in the results (\ref{resultp4}) and (\ref{resultp6}).

The argument for the singlet current coupling to mesons is just as true in
three- as in two-flavour ChPT. However here one needs to use the
three-flavour singlet current, $j_{I3}^\mu$, instead. Again denoting with
a tilde the contributions from loop diagrams involving
only lowest order vertices or NLO vertices not from the WZW term, we have
(after using isospin) two relations similar to (\ref{relprimeNf2})
\begin{align}
\label{relprimeNf3}
\pPi_{U(U+D+S)} =\,& \pPi_{U}+\pPi_{UD}+\pPi_{US} = 0\,,
\nonumber\\
\pPi_{S(U+D+S)} =\,& 2\pPi_{US}+\pPi_S = 0\,.
\end{align}
Note that in this subsection we talk about the three-flavour ChPT
expressions. Inspection of the expressions in
(\ref{resultp4}) and (\ref{resultp6}) show that the relations
(\ref{relprimeNf3}) are satisfied.
Note that the relation (\ref{relprimeNf3}) if written for $\hPi$ has corrections
at order $p^8$.

In general we can write using (\ref{relprimeNf3})
\begin{equation}
\label{su3relation}
\frac{\pPi_{UD}}{\pPi_{\pip}} = -\frac{1}{2} - \frac{\pPi_{US}}{2\pPi_{\pip}}\,.
\end{equation}
This indicates that corrections to the $-1/2$ are expected to be small
due to the strange quark being much heavier than the up and down quarks.

The second relation in (\ref{relprimeNf3}) allows a relation involving
two-point functions with the strange quark current.

Note that a consequence of (\ref{relprimeNf3}) in the equal mass limit is
\begin{equation}
m_u=m_d=m_s~\Longrightarrow~
\frac{\pPi_{UD}}{\pPi_{\pip}} = -\frac{1}{3}\,.
\end{equation}
In this case the disconnected contribution to the electromagnetic two-point
function vanishes identically since the charge matrix is traceless. 

\section{Estimate of the ratio of disconnected to connected}
\label{estimatedisconnected}

In order to estimate the ratio of disconnected to connected contributions in
ChPT the inputs that appear must be determined.
For the plots shown below we use
\begin{align}
F_\pi =& 92.2~\mathrm{MeV} &
m_\pi =& 135~\mathrm{MeV} &
m_K =& 495~\mathrm{MeV}
\nonumber\\
L_9^r =&0.00593 &
\mu =& 770~\mathrm{MeV}
\end{align}
The values for the decay constant and masses are standard ones.
The values for the $L_i^r$ were recently reviewed in
\cite{Bijnens:2014lea} and we have taken the values for
$L_9^r$ from \cite{Bijnens:2002hp} 
quoted in \cite{Bijnens:2014lea}.

If we only consider $\hPi$, the only other LECs we need
are $C_{93}^r$ and $D_3^r$.
As first suggested in \cite{Ecker:1988te} LECs are expected to be saturated
by resonances. For $C_{93}^r$ and $D_3^r$ the main contribution
will be from the vector
resonance multiplet. Here a nonet approach typically works well
and that would suggest that $D_3^r\approx 0$. We will set it to zero
in our estimates. The value for $C_{93}^r$ was first determined using
resonance saturation in \cite{Amoros:1999dp} with a value of
\begin{align}
C_{93}^r = -1.4~10^{-4}
\end{align}
If we use resonance saturation for the nonet and the constraints from
short-distance as used in \cite{Ecker:1989yg} we obtain for the two-point
function
\begin{align}
\label{piVMD}
\Pi_{\pi^+}^\mathrm{VMD}(q^2) = \frac{4 F_\pi^2}{m_V^2-q^2}\,.
\end{align}
Assuming that the pure LEC parts reproduce (\ref{piVMD}), leads to the value
\begin{align}
C_{93}^r = -1.02~10^{-4}
\end{align}
with $m_V=770$~MeV. Finally fitting the expression for $\Pi_{\pi^+}$
to a phenomenological form of the two-point function \cite{Golterman:2014ksa}
gives
\begin{align}
C_{93}^r = -1.33~10^{-4}
\end{align}
The three values are in reasonable agreement.
The size can be compared to other vector meson dominated combinations
of LECs, e.g. 
$C_{88}^r-C_{90}^r = -0.55~10^{-4}$  \cite{Bijnens:2002hp}, which is of the
same magnitude. In the numerical results we will use the full expression
(\ref{piVMD}) for the contribution from higher order LECs rather
than just the terms with $C_{93}^r$.

\begin{figure}[t]
\begin{minipage}{0.45\textwidth}
\includegraphics[width=0.99\textwidth]{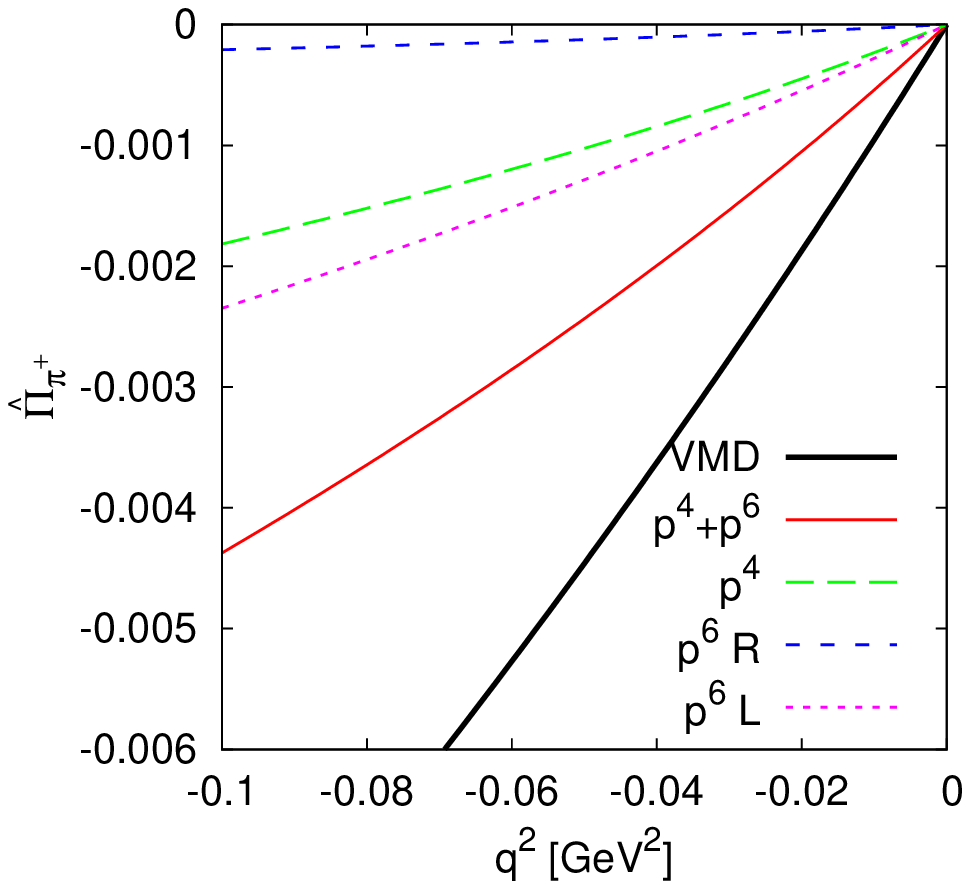}
\caption{The subtracted two-point function $\hPi_{\pi^+}(q^2)$ or
the connected part. Plotted are the $p^4$ contribution
of (\ref{resultp4}) labeled $p^4$ and the three parts of the higher
order contribution: the pure two-loop contribution labeled $p^6~R$,
the $p^6$ contribution from one-loop graphs labeled $p^6~L$
and the pure LEC contribution as modeled by (\ref{piVMD}) labeled VMD.}
\label{figpip}
\end{minipage}
~~~
\begin{minipage}{0.45\textwidth}
\includegraphics[width=0.99\textwidth]{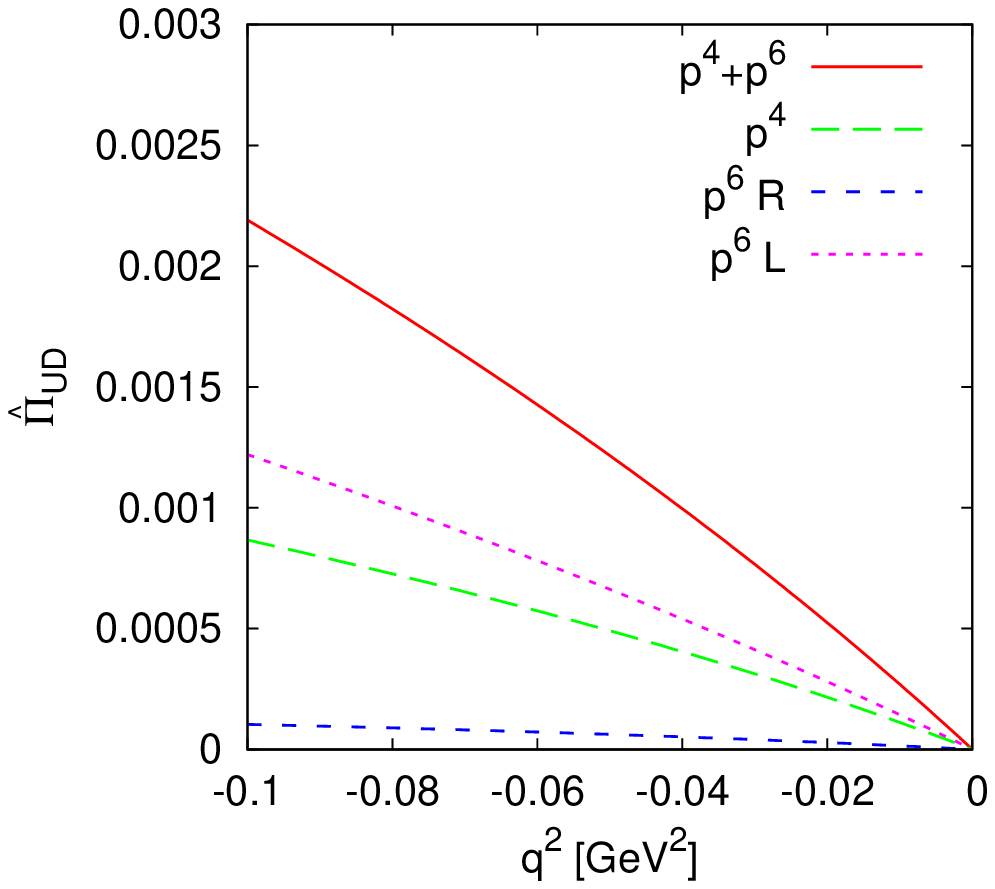}
\caption{The subtracted two-point function $\hPi_{UD}(q^2)$ or
the disconnected part. Plotted are the $p^4$ contribution
of (\ref{resultp4}) labeled $p^4$ and the two non-zero parts of the higher
order contribution: the pure two-loop contribution labeled $p^6~R$
and
the $p^6$ contribution from one-loop graphs labeled $p^6~L$.
The the pure LEC contribution is estimated to be zero here.}
\label{figpiUD}
\end{minipage}
\end{figure}

In Fig.~\ref{figpip} we have plotted the different contributions to
$\hPi_{\pi^+}$. This is what is usually called the connected contribution.
As we see, the contribution from higher order LECs,
as modeled by (\ref{piVMD}), is, as
expected, dominant. The full result for $\hPi$ is the sum of the VMD
and the $p^4+p^6$ lines. We see that the pure two-loop contribution
is small compared to the one-loop contribution but there is a large contribution
at order $p^6$ from the one-loop diagrams involving $L_i^r$.

In Fig.~\ref{figpiUD} we have plotted the same contributions but now for
$\hPi_{UD}$ or the contribution from disconnected diagrams.
Note that the scale is exactly half that of Fig.~\ref{figpip}.
The contributions are very close to $-1/2$ times those of Fig.~\ref{figpip}
except for the pure LEC contribution which is here estimated to be zero.

\begin{figure}[tbp]
\begin{center}
\begin{minipage}{0.45\textwidth}
\includegraphics[width=0.99\textwidth]{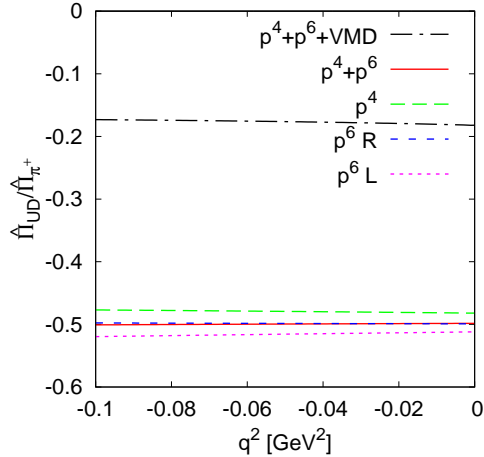}
\end{minipage}
\end{center}
\caption{The ratio of the subtracted two-point functions
$\hPi_{UD}(q^2)/\hPi_{\pi^+}(q^2)$ or ratio of 
the disconnected to the connected part. Plotted are the $p^4$ contribution
of (\ref{resultp4}) labeled $p^4$, the parts of the higher
order contribution: the pure two-loop contribution labeled $p^6~R$
and
the $p^6$ contribution from one-loop graphs labeled $p^6~L$
as well as their sum.
The ratio of the pure LEC contribution is estimated to be zero.
The ratio for all contributions summed is the dash-dotted line.}
\label{figpiratio}
\end{figure}

How well do the estimates of the ratio now hold up.
The ratio of disconnected to connected is plotted in Fig.~\ref{figpiratio}.
We see that the contribution at order $p^4$ has a ratio very close to $-1/2$
and the same goes for all loop contributions at order $p^6$. The effects
of kaon loops is thus rather small. The deviation from
$-1/2$ is driven by the estimate of the pure LEC contribution.
Using the VMD estimate (\ref{piVMD}) we end up with a ratio of about $-0.18$
for the range plotted. Taking into account (\ref{PiEM2}) we get
an expected ratio for the disconnected to connected contribution
to the light quark electromagnetic two-point function
$\hPi_{EM2}$ of about $-3.5\%$. If we had used the other estimates
for $C_{93}^r$ (and assumed a similar ratio for higher orders)
the number would have been about $-3\%$.

An analysis using only the pion contributions, so no contribution from
intermediate kaon states, would give essentially the same result.

\section{Estimate of the strange quark contributions}
\label{estimatestrange}

The numerical results in the previous section included the contribution
from kaons but only via the electromagnetic couplings to up and down quarks.
In this section we provide an estimate for the contribution when
including the photon coupling to strange quarks,
i.e. we add the terms coming from $\Pi_{US}$ and $\Pi_S$ in (\ref{PiEM}).
 
The loop contributions satisfy the relations shown in
(\ref{relprimeNf3}) with corrections starting earliest at $p^8$.
Alternatively we can write the first relation as
\begin{align}
\pPi_{\pi^+}+2\pPi_{UD}+\pPi_{US}=0\,,
\end{align}
this, together with the ratios shown in Fig.~\ref{figpiratio} and
the second relation in (\ref{relprimeNf3}), shows that we can expect
the extra contributions
to be quite small with the possible exception of the pure LEC contribution.

The pure LEC contribution is estimated to only apply to the connected
part and so contributes only to $\Pi_S$. Given that the $\phi$ mass
is significantly larger than the $\rho$-mass we will for that part
need to include this difference. A first estimate is simply by using
(\ref{piVMD}) with $m_V$ now the $\phi$-mass of $m_\phi=1020$~MeV.
We will call this VMD$\phi$ in the remainder.

The estimate we include for $\Pi_S$ includes both connected and disconnected
contributions. We would need to go to partially quenched ChPT to obtain
that split-up generalizing the methods of \cite{DellaMorte:2010aq}.

\begin{figure}[tbp]
\begin{center}
\begin{minipage}{0.45\textwidth}
\includegraphics[width=0.99\textwidth]{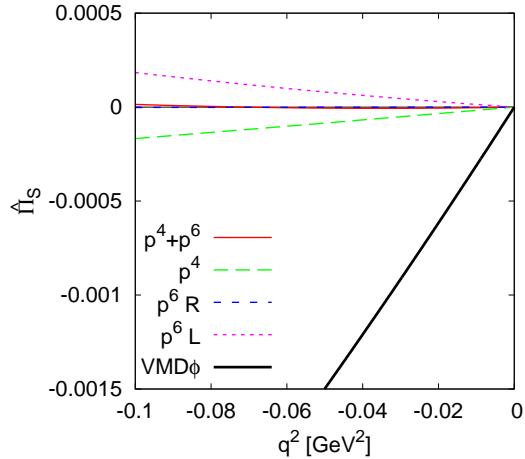}
\end{minipage}
\end{center}
\caption{The subtracted two-point function
$\hPi_{S}(q^2)$.
Plotted are the $p^4$ contribution
of (\ref{resultp4}) labeled $p^4$, the parts of the higher
order contribution: the pure two-loop contribution labeled $p^6~R$
and the $p^6$ contribution from one-loop graphs labeled $p^6~L$
as well as their sum.
The pure LEC contribution is estimated by (\ref{piVMD}) with the mass
of the $\phi$.}
\label{figpiS}
\end{figure}

Fig.~\ref{figpiS} shows the different contributions to $\hPi_S$.
We did not plot $\hPi_{US}$ since the relations
(\ref{relprimeNf3}) imply that the $p^4$, $p^6 L$ and $p^6 R$ are exactly $-1/2$
the contributions for $\hPi_S$ and in our estimate
the pure LEC part for $\hPi_{US}$ vanishes.
The contributions are much smaller than those of the connected light quark
contribution shown in Fig.~\ref{figpip}. One remarkable effect is the very
strong cancellation between the $p^4$ and $p^6$ effects give an
almost zero loop contribution. This means that vector meson dominance in the
coupling is even more clear in this case than for the lighter quarks.

\section{Comparison with lattice and other data}
\label{latticecomparison}

For comparing with lattice and phenomenological data we can use
the Taylor expansion around $q^2=0$ from our expressions and
the same coefficients evaluated from experimental data or via the
time moment analysis on the lattice \cite{Chakraborty:2014mwa}.

We expand the functions as
\begin{align}
\hPi(q^2) = \Pi_1 q^2 - \Pi_2 q^4 + \cdots
\end{align}
The signs follow from the fact that the lattice expansion is defined in terms
of $Q^2=-q^2$ and the usual lattice convention for $\Pi$ has the opposite sign
of ours. 
The coefficients, obtained by fitting an eight-order polynomial
to the ranges shown in the plots, are given in Table~\ref{tabPii}.

\cite{Benayoun:2016krn} is from an analysis of experimental data.
\cite{MiuraLattice2016} are preliminary numbers from the BMW collaboration
and we have removed the charm quark contribution from their numbers.
These numbers are not corrected for finite volume.
For \cite{Chakraborty:2014mwa,Chakraborty:2016mwy} we have taken the numbers
from their configuration 8, which has physical pion masses
and multiplied by $9/5$ for the latter to obtain $\Pi_{\pi^+}$.
\begin{table}[tbh]
\begin{center}
\newcommand{\m}{\phantom{-}}
\begin{tabular}{ccll}
\hline
Reference & $\Pi_A$ & $\Pi_1$ (GeV$^{-2}$) & $\Pi_2$ (GeV$^{-4}$)\\ 
\hline
$\Pi^{VMD}$ & $\hPi_{\pi^+}$ &  $\m0.0967$ & $-0.163$ \\
$p^4$ & $\hPi_{\pi^+}$       & $\m0.0240$ & $-0.091$ \\
$p^6~R$ & $\hPi_{\pi^+}$     & $\m0.0031$ & $-0.014$ \\
$p^6~L$ & $\hPi_{\pi^+}$     & $\m0.0286$ & $-0.067$ \\
sum & $\hPi_{\pi^+}$         & $\m0.152$ & $-0.336$\\
\cite{MiuraLattice2016} & $\hPi_{\pi^+}$ &$0.1657(16)(18)$ & $-0.297(10)(05)$\\
\cite{Chakraborty:2016mwy} & $\hPi_{\pi^+}$ & $0.1460(22)$ & $-0.2228(65)$ \\ 
\hline
$p^4$ & $\hPi_{UD}$ & $-0.0116$   & $\m0.045$ \\
$p^6~R$ & $\hPi_{UD}$ & $-0.0015$ & $\m0.007$ \\
$p^6~L$ & $\hPi_{UD}$ & $-0.0146$ & $\m0.032$ \\
sum & $\hPi_{UD}$ & $-0.0278$  & $\m0.085$\\
\cite{MiuraLattice2016} & $\hPi_{UD}$ &$-0.015(2)(1)$ & $\m0.046(10)(04)$\\
\hline
$\Pi^{VMD\phi}$ & $\hPi_{S}$ & $\m0.0314$ & $-0.030$ \\
$p^4$ & $\hPi_{S}$          & $\m0.0017$ & $-0.001$ \\
$p^6~R$ & $\hPi_{S}$        & $\m0.0000$ & $\m0.000$ \\
$p^6~L$ & $\hPi_{S}$        & $-0.0013$ &  $-0.005$ \\
sum & $\hPi_{S}$            & $\m0.0318$ & $-0.035$\\
\cite{MiuraLattice2016} & $\hPi_{S}$ &$\m0.0657(1)(2)$ & $-0.0532(1)(3)$\\
\cite{Chakraborty:2014mwa} & $\hPi_{S}$ & $\m0.06625(74)$ & $-0.0526(11)$\\
\hline
our result              & $\hPi_{EM}$ & $\m0.0852$    & $-0.182$ \\
\cite{Benayoun:2016krn} & $\hPi_{EM}$  & $\m0.0990(7)$ & $-0.206(2)$ \\
\cite{MiuraLattice2016} & $\hPi_{EM}$ & $\m0.0972(2)(1)$ & $-0.166(6)(3)$\\
\hline
\end{tabular}
\end{center}
\caption{The Taylor expansion coefficients of $\hPi$
of \cite{Chakraborty:2014mwa,Benayoun:2016krn,MiuraLattice2016,Chakraborty:2016mwy} and a comparison
with our estimates.}
\label{tabPii}
\end{table}
Our estimates are in reasonable agreement for the connected contribution.
For the disconnected contribution, our results are higher but of a similar
order.

There have been many more studies of the muon $g-2$ on the lattice
and in particular a number of studies of the disconnected part.
However, their results are often not presented in a form that we can easily
compare to. From our numbers above we expect the disconnected contribution
to be a few \% and of the opposite sign of the connected contribution.
\cite{Chakraborty:2015ugp} finds $-0.15(5)\%$, much smaller than we expect,
\cite{Blum:2015you} finds about $-1.5\%$ which is below but of the same
order as our estimate.

The same comment applies to studies of the strange contribution, e.g. 
\cite{Blum:2016xpd} finds a contribution of about 7\% of the light
connected contribution which is in reasonable agreement with our estimate.

\section{Summary and conclusions}
\label{conclusions}

We have calculated in two- and three-flavour ChPT all the neutral two-point
functions in the isospin limit including the singlet vector current.
We have extended the ratio of $-1/2$ (or $-1/10$ for the electromagnetic
current) of \cite{DellaMorte:2010aq} to a large part of the higher order
loop corrections. We used the nonet estimates of LECs to set the new constants
for the singlet current equal to zero and then provided numerical estimates
for the disconnected and strange quark contributions.

We find that the disconnected contribution is negative and a few \% of the
connected contribution, the main uncertainty being the new LECs which we
estimated to be zero. A similar estimate for the strange quark contribution
has a large cancellation between $p^4$ and $p^6$ leaving our rather uncertain
estimate of the LECs involved as the main contribution.

\section*{Acknowledgements}

This work is supported in part by the Swedish Research Council grants
contract numbers 621-2013-4287 and 2015-04089 and by
the European Research Council (ERC) under the European Union's Horizon 2020
research and innovation programme (grant agreement No 668679).

\end{document}